\documentclass[conference]{IEEEtran}

\usepackage[final]{graphicx}
\usepackage{times}
\usepackage{amsfonts}
\usepackage{epsfig}
\usepackage{amsbsy,amssymb}
\usepackage{amsmath}
\usepackage{citesort}
\usepackage{subfigure}
\usepackage{inputenc}

\newcommand{\beq}{\begin{equation}}
\newcommand{\eeq}{\end{equation}}
\newcommand{\ben}{\begin{enumerate}}
\newcommand{\een}{\end{enumerate}}
\newcommand{\bit}{\begin{itemize}}
\newcommand{\eit}{\end{itemize}}

\DeclareMathOperator{\Var}{Var}

\begin{document}


\title{A Novel Channel Estimation based on Spread Pilots for Terrestrial Digital Video Broadcasting}
\author{Oudomsack Pierre Pasquero, Matthieu Crussi\`ere, Youssef Nasser, Jean-Fran\c{c}ois H\'elard\\
{Institute of Electronics and Telecommunications of Rennes (IETR)} \\
{INSA Rennes, 20, avenue des Buttes de Coesmes, 35000 Rennes, France}\\
{E-mail: oudomsack.pasquero@ens.insa-rennes.fr, \{first name. last name\}@insa-rennes.fr}}
\maketitle

\begin{abstract}
In this paper, we propose a novel channel estimation technique based on spread pilots for digital video broadcasting. This technique consists in adding a linear precoding function before the OFDM modulation and dedicating one of the precoding sequence to transmit the pilot symbols for the channel estimation. The merits of this technique are its simplicity, its flexibility, and the gains in terms of spectral efficiency and useful bit rate obtained compared to the classical pilot based estimation schemes used in DVB standards. The performance evaluated over realistic channel models, shows the efficiency of this technique which turns out to be a promising channel estimation technique for the future terrestrial video broadcasting systems.

\end{abstract}

\section{INTRODUCTION}
Since its inception in 1997, Digital Video Broadcasting Terrestrial (DVB-T) standard \cite{DVBT} has fully responded to the objectives of its designers, delivering wireless digital TV services in almost every continent. However, the emergence of new consumer usages is leading the DVB community to think about a second generation of DVB-T. In that context, a European project called Broadcasting for the 21st Century (B21C) has recently been launched \cite{B21C}. It constitutes a contribution task force to the reflections engaged within DVB-T2. The study proposed in this paper has been carried out within the framework of B21C. 

Orthogonal Frequency Division Multiplexing (OFDM) has been perceived as one of the most effective transmission schemes for the frequency-selective fading channels. Indeed, by implementing Inverse Fast Fourier Transform (IFFT) at the transmitter and Fast Fourier Transform (FFT) at the receiver, OFDM splits the single channel into multiple, parallel intersymbol interference (ISI) free subchannels. Therefore, each subchannel, also called subcarrier, can be easily equalized by only one coefficient.

To equalize the OFDM signal, the receiver needs to estimate the channel frequency response for each subcarrier. In the DVB-T standard, one subcarrier over twelve is used as pilot, and interpolating filtering techniques are applied to obtain the channel response for any subcarrier. However, these pilots dramatically reduce the spectral efficiency of the system. The originality of this work is to propose a new channel estimation approach based on a linear precoded (LP) multicarrier waveform. The basic idea consists in using a two-dimensional (2D) linear precoding matrix before the OFDM modulation, and to dedicate one of the precoding sequence, also called spreading sequence, to transmit a so-called spread pilot information for channel estimation \cite{Cariou_EL07}. This technique is shown to provide a good flexibility owing to the 2D LP function, better spectral efficiencies and useful bit rates compared to those of DVB-T. In addition, note that the precoding component can also be exploited to reduce the peak-to-average ratio (PAPR) of the multicarrier system \cite{Nobilet_TransOnTelecom02}, or to perform frequency synchronisation.

This paper is organized as follows. After a reminder of the 2D LP OFDM concept, we describe in \mbox{section II} the principle of the channel estimation using the spread pilots. The theoretical performance of the estimator is discussed in \mbox{section III}. Then, simulation results in term of bit error rate (BER) and comparison with the DVB-T system are given in \mbox{section IV}. Concluding remarks are given in \mbox{section V}.

\section{SYSTEM DESCRIPTION}

\subsection{2D LP OFDM}
\begin{figure} [h]
	\begin{center}
		\includegraphics[width=1 \linewidth]{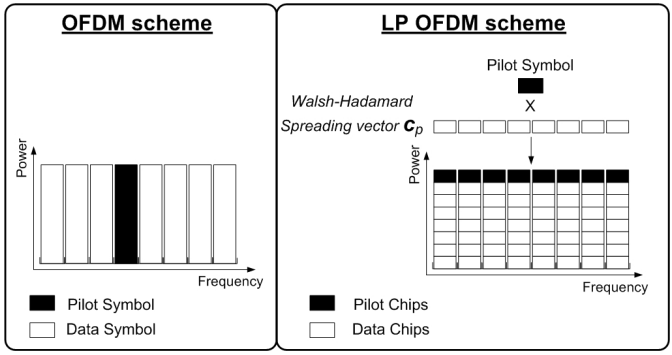}
		\caption{Example of 1D LP OFDM in frequency domain compared to OFDM}
		\label{MCSS}
	\end{center}
\end{figure}

\begin{figure} [h]
\begin{center}
		\includegraphics[width=0.8 \linewidth]{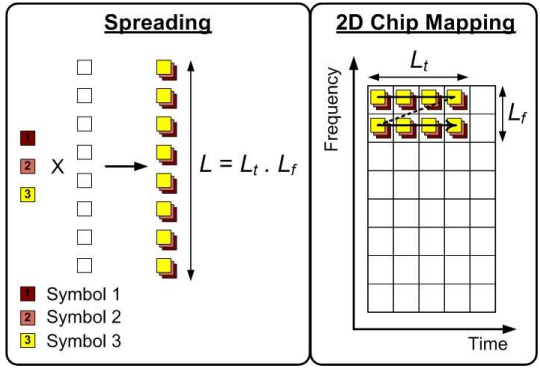}
  	\includegraphics[width=0.9 \linewidth]{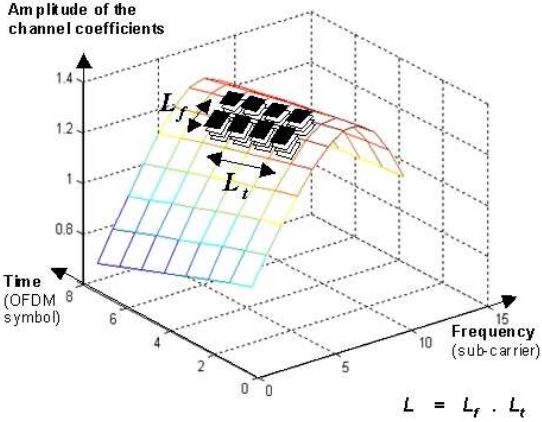}
\caption{Principle of a 2D chip mapping following a zigzag in time}
\label{2DChipMapping}
\end{center}
\end{figure}

Precoding an OFDM signal consists in spreading each complex symbol $x_{i}$ before the OFDM modulation, over $L = 2^n$ chips, with $n \in \mathbb{N}$, using Walsh-Hadamard (WH) sequences $\textbf{c}_{i}=\left[ c_{i1} \dots c_{ij} \dots c_{iL} \right]^{T}$. We assume the data symbols have zero mean and unit variance, and the spreading sequences are normalized, \textit{i.e.} $\left|{c}_{ij}\right|^{2} = \frac{1}{L}$. A superposition of the data chips and the pilot chips, as illustrated in Fig.\ref{MCSS} and Fig.\ref{2DChipMapping}, generates the chip stream:
\begin{equation}
\textbf{s} = \sum^{L}_{i=1 \atop i\neq{p}} \textbf{c}_{i} x_{i} + \textbf{c}_{p} \sqrt{B} x_{p}
\end{equation}
where $x_{p}$ corresponds to the pilot symbol, and $B$ represents a factor related to an eventual power boost applied on it. 

The chips obtained are mapped over a subset of $L=L_{t}.L_{f}$ subcarriers, with $L_{t}$ and $L_{f}$ the time and frequency spreading factors respectively. The first $L_t$ chips are allocated in the time direction. The next blocks of $L_t$ chips are allocated identically on adjacent subcarriers as illustrated in Fig. \ref{2DChipMapping}. Therefore, the 2D chip mapping follows a zigzag in time. Note that the chip mapping can also follow a zigzag in frequency, a snake in time or a snake in frequency \cite{Chapalain_PIMRC05}. 
\begin{figure} [h!]
	\begin{center}
		\includegraphics[width=1\linewidth]{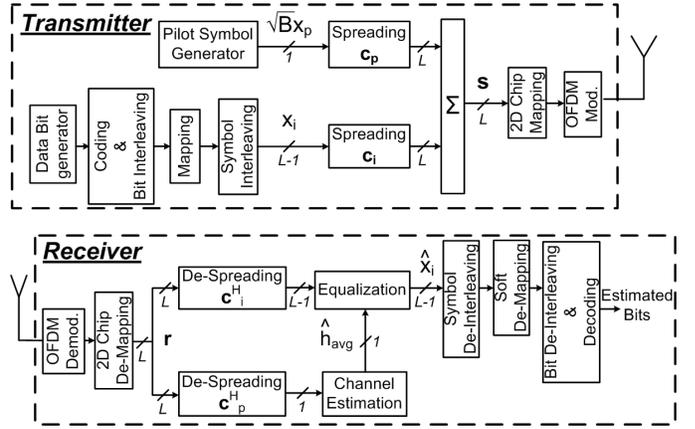}
		\caption{2D LP OFDM transmitter and receiver based on spread pilot channel estimation technique}
		\label{TX_RX}
	\end{center}
\end{figure} 
\subsection{Principle of channel estimation with spread pilots}
Let $\textbf{h}=\left[h_{1} \dots h_{j} \dots h_{L}\right]^{T}$ be the vector of the channel coefficients associated to the $L$ subcarriers on which the chips have been mapped. After OFDM demodulation, the received signal vector $\textbf{r}$ writes: 
\begin{align}
\textbf{r} &= \textbf{h} \circ \textbf{s} + \textbf{n} \nonumber \\
&= \textbf{h} \circ \left( \textbf{c}_{p} \sqrt{B} x_{p} + \sum^{L}_{i=1 \atop i\neq{p}} \textbf{c}_{i} x_{i} \right) + \textbf{n}
\end{align}
where $\circ$ denotes the element-wise vector multiplication and vector $\textbf{n}=\left[n_1 \dots n_j \dots n_L \right]^{T}$ gathers additive white Gaussian noise (AWGN) components having zero mean and variance $\sigma_{n}^{2}$.

Fig.~\ref{TX_RX} depicts the transmitter and receiver schemes. At the reception, the de-spreading function is processed before equalization. Therefore, one single channel coefficient $\widehat{h}_{avg}$ is estimated for each subset of subcarriers and is used to equalize the $(L-1)$ received data symbols mapped on that subset. This estimated channel coefficient is obtained by de-spreading the received signal $\textbf{r}$ using the pilot spreading sequence $\textbf{c}^{H}_{p}$, and dividing by the pilot symbol $x'_p=\sqrt{B}x_p$:
\small
\begin{align} 
\widehat{h}_{avg} &= \frac{1}{x'_p} \textbf{c}^{H}_{p} \textbf{r} \nonumber \\
&= \frac{1}{x'_p} \left( \sum_{j=1}^{L} h_j \left(c_{pj}^2 x'_p \right) + \sum_{i=1 \atop i\neq p}^{L} \sum_{j=1}^{L}  c_{pj} h_j \left( c_{ij} x_i \right) + \sum_{j=1}^{L} c_{pj} n_j  \right)  \nonumber \\
&= \frac{1}{L} \sum_{j=1}^{L} h_j + \frac{1}{x'_p} \left(\sum_{i=1 \atop i\neq p}^{L} x_i \sum_{j=1}^{L} c_{pj} h_j c_{ij} + \sum_{j=1}^{L} c_{pj} n_j \right) \nonumber \\
\end{align} 
\normalsize
Let us define the average channel coefficient hang of the considered subset of subcarriers $h_{avg}=\frac{1}{L} \sum_{j=1}^{L} h_j$ and assume $\epsilon_j = h_j - h_{avg}$. The estimated channel coefficient can now be separated into the average channel coefficient $h_{avg}$, a self-interference (SI) term and a noise term $n'$.
\begin{align} 
\widehat{h}_{avg} &= h_{avg} + \underbrace{\sum_{i=1 \atop i\neq p}^{L} \frac{x_i}{x'_p} \sum_{j=1}^{L} c_{pj} \epsilon_j c_{ij}}_{\textrm{SI}} + \underbrace{\sum_{j=1}^{L} \frac{c_{pj} n_j}{x'_p}}_{n'}
\end{align}
The SI term is due to the variation of the channel coefficients over the subset of subcarriers which causes the loss of orthogonality between the spreading sequences. In the sequel, we propose to analyse its variance. 

\begin{figure*} [t!!!!]
	\begin{center}
			\subfigure[MSE simulation results for F1 channel]{%
				\label{Fig:MSE}
				\includegraphics[width=0.45\linewidth]{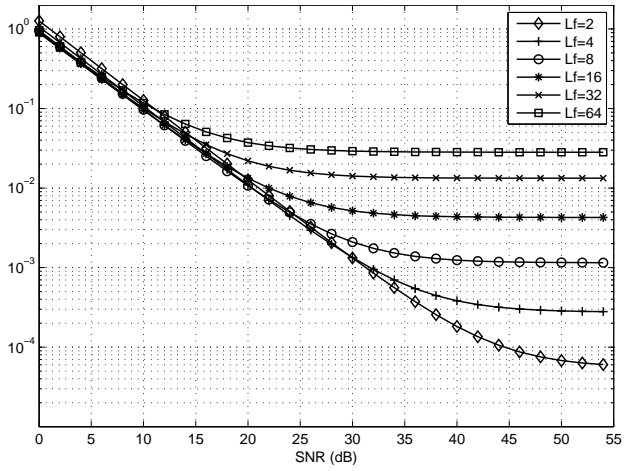}}
  		\subfigure[Theoretical F1 weighted channel variance]{%
  			\label{Fig:var}
  			\includegraphics[width=0.45\linewidth]{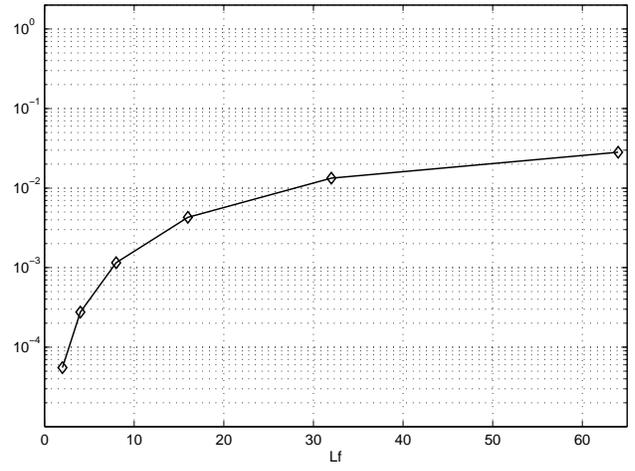}}
		\caption{Channel estimator performance under F1 channel, $B=1$, $L_t=1$}
		\label{Fig:MSEvar}
	\end{center}
\end{figure*}

\section{THEORETICAL PERFORMANCE OF THE ESTIMATOR}
In order to analyse the theoretical performance of the estimate $\widehat{h}_{avg}$, let us compute its expectation E\{$\widehat{h}_{avg}$\} and its mean square error (MSE).
\small
\begin{align} \label{NotBiased1}
\textrm{E}\left\{\widehat{h}_{avg}\right\} &= h_{avg} + \textrm{E}\left\{\textrm{SI}\right\} + \textrm{E}\left\{n'\right\}  \nonumber \\
&= h_{avg} +  \sum_{i=1 \atop i\neq p}^{L} \frac{\textrm{E}\left\{ x_i \right\}}{x'_p}  \sum_{j=1}^{L}  \textrm{E}\left\{c_{pj} \epsilon_j c_{ij}\right\} + \sum_{j=1}^{L} \frac{\textrm{E}\left\{c_{pj} n_j\right\}}{x'_p} 
\end{align}
\normalsize
Since the data symbols $x_i$ and the noise components $n_j$ have zero mean, the SI and the noise contributions are null. Hence, (\ref{NotBiased1}) can simply be rewritten:
\begin{align} \label{NotBiased2}
\textrm{E}\left\{\widehat{h}_{avg}\right\} &= h_{avg}
\end{align}
Equation (\ref{NotBiased2}) proves that $\widehat{h}_{avg}$ is not biased. Consequently, the MSE of the estimator is equal to its variance:
\begin{align} \label{MSE}
\textrm{MSE} &= \textrm{E}\left\{\left|\widehat{h}_{avg}-\textrm{E}\left\{\widehat{h}_{avg}\right\}\right|^{2}\right\} \nonumber \\
&= \textrm{E}\left\{\left|\textrm{SI}\right|^{2}\right\} + \textrm{E}\left\{\left| n' \right|^{2}\right\}
\end{align}
First of all, let us analyse the SI variance:
\begin{align} \label{SI}
\Var\left\{\textrm{SI}\right\} &= \frac{1}{B} \left( \sum_{i=1 \atop i\neq{p}}^{L} \textrm{E}\left\{ \left|x_i\right|^{2} \right\} \sum_{j=1}^{L} \textrm{E}\left\{ \left| c_{pj} c_{ij} \right|^{2} \right\} \textrm{E}\left\{ \left|\epsilon_j\right|^{2} \right\}  \right)  \nonumber \\
&= \frac{1}{B} \left( \left(L-1\right) \frac{1}{L^{2}} \sum_{j=1}^{L} \textrm{E}\left\{ \left| \epsilon_j \right|^{2} \right\} \right) \nonumber \\
&= \frac{1}{B} \left( \frac{\left(L-1\right)}{L} \sigma^{2}_{h} \right) 
\end{align}
where $\sigma^{2}_{h} = \frac{1}{L} \sum_{j=1}^{L} \textrm{E}\left\{ \left| \epsilon_j \right|^{2} \right\}$ corresponds to the channel variance on the studied subset of subcarriers. Now, let us compute the noise variance:
\begin{align} \label{noise}
\Var\left\{n'\right\} &= \frac{1}{B} \sum_{j=1}^{L} \textrm{E}\left\{ \left| c_{pj} \right|^{2} \right\} \textrm{E}\left\{ \left| n_j \right|^{2} \right\} \nonumber \\
&= \frac{1}{B} \sigma^{2}_{n}
\end{align}
Finally, by combining the SI variance (\ref{SI}) and the noise variance (\ref{noise}), the MSE of the estimator (\ref{MSE}) can be expressed as:
\begin{align} \label{MSE2}
\textrm{MSE} &= \frac{1}{B} \left( \frac{\left(L-1\right)}{L} \sigma^{2}_{h} + \sigma^{2}_{n} \right)
\end{align}
We deduce that the MSE is proportional to the channel variance on the specific subset of subcarriers and is attenuated by the boost factor $B$. One can actually check that if the channel is flat over a subset of subcarriers, i.e. $\sigma^{2}_{h}=0$, then the SI is null. Therefore, it is important to optimize the time and frequency spreading lengths, $L_t$ and $L_f$, according the channel characteristics. In addition, the boost factor $B$ has to be chosen adequatly.

\section{SIMULATION RESULTS}
In this section, we analyze the performance of the proposed channel estimation scheme under F1 and P1 channel models \cite{DVBT}. These channel models are modelized without any Doppler effect. They are specified for fixed outdoor rooftop antenna reception conditions. Table \ref{SimParam} gives the simulation parameters and the useful bit rates of the DVB-T system and the proposed 2D LP OFDM system. In the proposed system, only one spread pilot symbol is used over $L\geq32$, whereas the DVB-T system uses one pilot subcarrier over twelve. Therefore, a gain in terms of spectral efficiency and useful bit rates is obtained compared to the DVB-T system. 
\begin{table} 
	\caption{Simulation Parameters and Useful Bit Rates}
	\begin{center}
		\begin{tabular}{|l|l|}
			\hline
			 Bandwidth & 8 MHz \\
			\hline
			 FFT size &  2048 samples \\
			\hline
			 Guard Interval size &  512 samples \\
			\hline
			 Constellations &   16QAM and 64QAM\\
			\hline
			 Polynomial code generator  &  $\left(133,171\right)_{o}$ \\
			\hline
			 Rate of convolutional code Rc &  3/4 and 5/6 \\
			\hline 
			\hline
			 Useful bit rates of DVB-T system 	& 14.93 Mbits/s	for 16QAM and Rc=3/4\\
			 																		& 24.88 Mbits/s	for 64QAM and Rc=5/6\\
			\hline
			 Useful bit rates of 2D LP OFDM			& 16.00 Mbits/s for $L=16$ \\
			 for 16QAM and Rc=3/4								& 16.53 Mbits/s for $L=32$ \\
																					& 16.80 Mbits/s for $L=64$ \\
			\hline
			 Useful bit rates of 2D LP OFDM			& 26.67 Mbits/s for $L=16$ \\
			 for 64QAM and Rc=5/6								& 27.55 Mbits/s for $L=32$ \\
																					& 27.99 Mbits/s for $L=64$ \\
			\hline
		\end{tabular}
	\end{center}
	\label{SimParam}
\end{table} 

Fig.\ref{Fig:MSEvar} depicts the estimator performance. Fig.\ref{Fig:MSE} gives the MSE of the channel estimation for different signal to noise ratios (SNR), while Fig.\ref{Fig:var} exhibits the weighted channel variance for different frequency spreading factors $L_{f}$. We note that beyond a given SNR, the MSE reaches a floor which is easily interpreted as being due to the channel variance $\sigma^{2}_{h}$, as expected from (\ref{MSE2}). This is confirmed by comparing Fig.\ref{Fig:MSE} and Fig.\ref{Fig:var}. Hence, in order to minimize the performance degradation, $L_f$ has to be chosen not too high. In the sequel, we set $L_f=2$ so that the subcarrier subsets do not exceed the coherence bandwidth of the channel.

\begin{figure} 
	\begin{center}
		\includegraphics[width=1\linewidth]{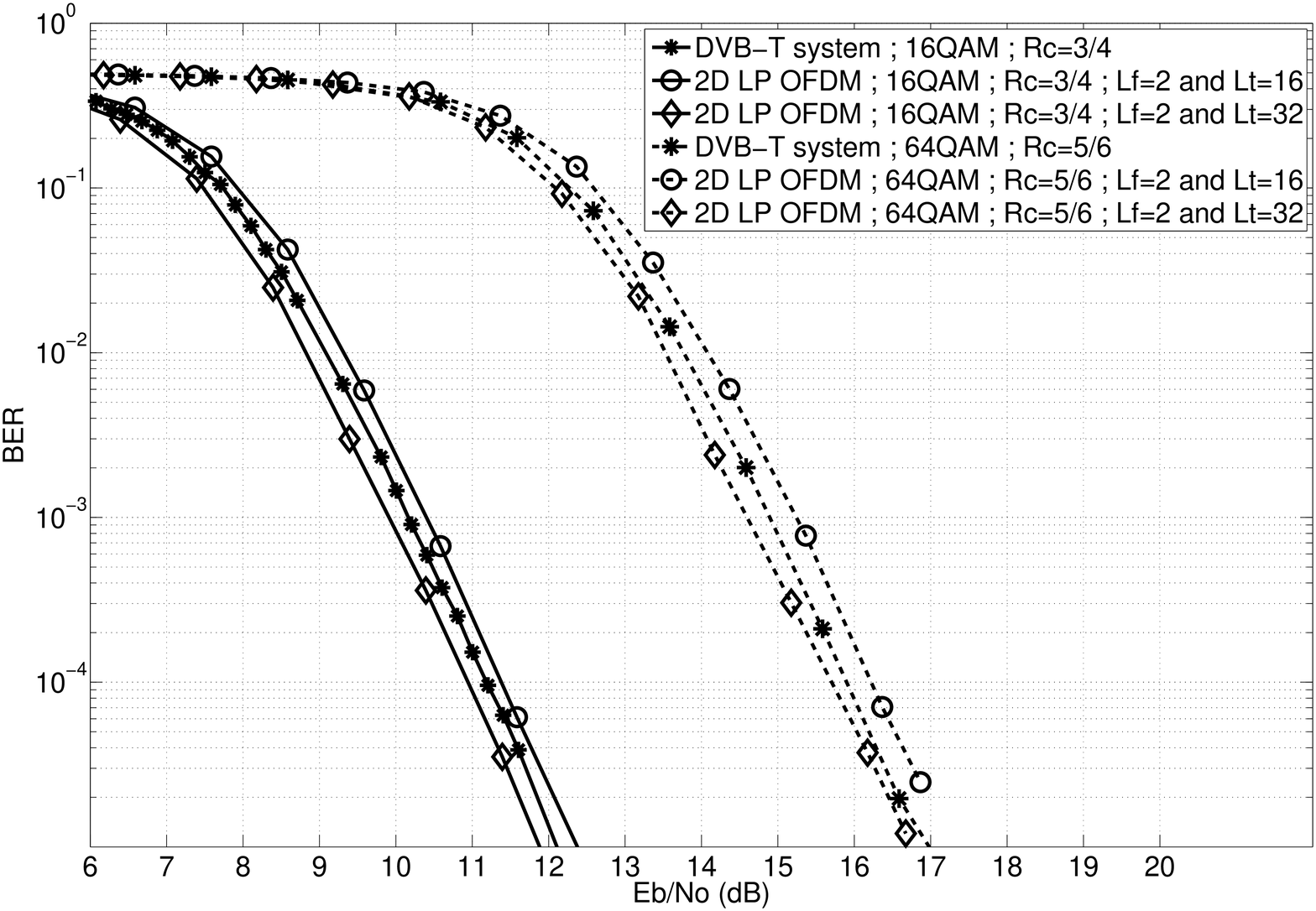}
		\caption{BER performance under F1 channel models}
		\label{ber_F1}
	\end{center}
\end{figure}

\begin{figure} 
	\begin{center}
		\includegraphics[width=1\linewidth]{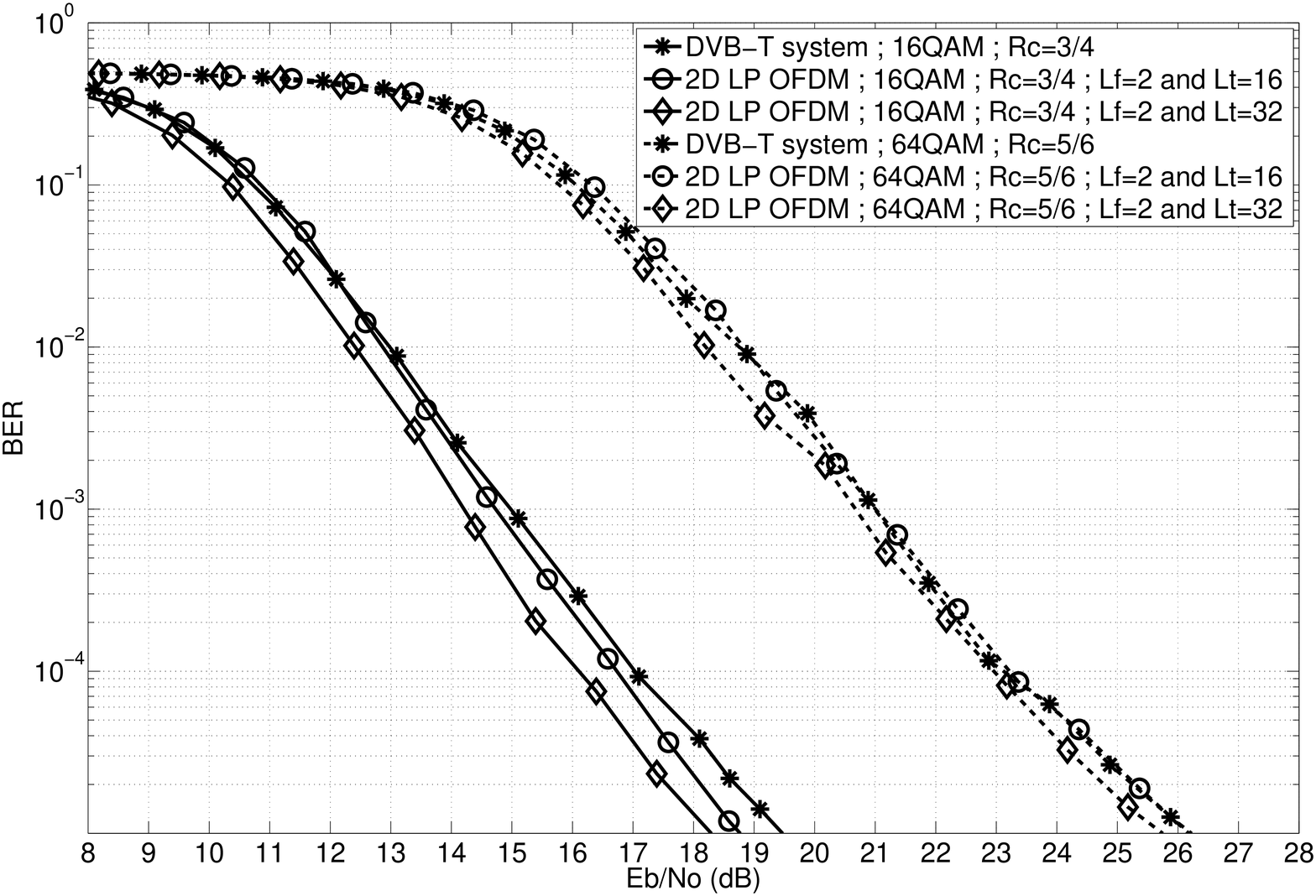}
		\caption{BER performance under P1 channel models}
		\label{ber_P1}
	\end{center}
\end{figure}

Fig.\ref{ber_F1} and Fig.\ref{ber_P1} give the BER measured at the output of the Viterbi decoder for the proposed precoded OFDM system using the spread pilot channel estimation scheme, for 16-QAM and 64-QAM, under F1 and P1 channel models respectively. The spreading component is applied in both time and frequency dimensions, for different values of $L_{t}$ keeping $L_f=2$. Note that the value of the boost factor $B$ has been optimized through simulation search in order to obtain the lowest BER for a given SNR. The performance of the DVB-T system with perfect channel estimation is given as reference, however taking into account the power loss due to the amount of energy spent for the transmission of the pilot subcarriers. It appears that the system performance is all the better than $L_t$ is high. The reason is the increase of the spectral efficiency due to the use of only one spread pilot symbol over $L$ symbols. For example, with $L=64$, \textit{i.e.} $L_t=32$, only one spread pilot symbol over $L=64$ is used whereas the DVB-T system uses one pilot subcarrier over twelve. Consequently, for high $L_t$ values, the proposed system can even outperform the classical DVB-T system with perfect channel estimation when the performance is given versus the $\frac{Eb}{No}$ ratio.

\section{CONCLUSION}
In this paper, we have proposed a novel and very simple channel estimation for DVB-T. This technique, referred to as spread pilot channel estimation, allows to reduce the overhead part dedicated to channel estimation. Therefore, a gain in terms spectral efficiency and useful bit rate is obtained. Moreover, this technique is very well suited to fixed reception conditions. Indeed, by choosing the highest possible time spreading factor allowed by memory constraints, and a reasonnable frequency spreading factor, the system performance outperforms the classical pilot-based DVB-T system. More generally, this estimation approach provides good flexibility since it can be optimized for different mobility scenarios choosing adequate time and frequency spreading factors.

\textit{Acknowledgement: This work was supported by the European project CELTIC B21C (``Broadcasting for the 21st Century'') and the French national project Mobile TV World.}

\bibliographystyle{IEEEtran}
\bibliography{ppasquero}

\begin{thebibliography}{1}
\providecommand{\url}[1]{#1}
\csname url@rmstyle\endcsname
\providecommand{\newblock}{\relax}
\providecommand{\bibinfo}[2]{#2}
\providecommand\BIBentrySTDinterwordspacing{\spaceskip=0pt\relax}
\providecommand\BIBentryALTinterwordstretchfactor{4}
\providecommand\BIBentryALTinterwordspacing{\spaceskip=\fontdimen2\font plus
\BIBentryALTinterwordstretchfactor\fontdimen3\font minus
  \fontdimen4\font\relax}
\providecommand\BIBforeignlanguage[2]{{%
\expandafter\ifx\csname l@#1\endcsname\relax
\typeout{** WARNING: IEEEtran.bst: No hyphenation pattern has been}%
\typeout{** loaded for the language `#1'. Using the pattern for}%
\typeout{** the default language instead.}%
\else
\language=\csname l@#1\endcsname
\fi
#2}}

\bibitem{DVBT}
{ETSI EN 300 744}, Tech. Rep.

\bibitem{B21C}
http://www.celtic initiative.org/Projects/{B21C}.

\bibitem{Cariou_EL07}
L.~Cariou and J.-F. H\'elard, ``Efficient mimo channel estimation for linear
  precoded ofdma uplink systems,'' \emph{Electronics Letters}, vol.~43, no.~18,
  pp. 986--988, 31 2007.

\bibitem{Nobilet_TransOnTelecom02}
S.~Nobilet, J.-F. H\'elard, and D.~Mottier, ``Spreading sequences for uplink
  and downlink mc-cdma systems: {PAPR} and {MAI} minimization,'' \emph{European
  Trans. on Telecommun.}, vol.~13, pp. 465--471, Oct. 2002.

\bibitem{Chapalain_PIMRC05}
N.~Chapalain, D.~Mottier, and D.~Castelain, ``Performance of uplink ss-mc-ma
  systems with frequency hopping and channel estimation based on spread
  pilots,'' \emph{Personal, Indoor and Mobile Radio Communications, 2005. PIMRC
  2005. IEEE 16th International Symposium on}, vol.~3, pp. 1515--1519 Vol. 3,
  Sept. 2005.

\end{thebibliography}




\end{document}